\documentclass[conference]{IEEEtran}
\IEEEoverridecommandlockouts
\usepackage{amssymb, amsmath, amsfonts}
\usepackage{algpseudocode,algorithm,algorithmicx}

\usepackage{graphicx}
\usepackage{graphics}
\usepackage{float}
\usepackage[compress,nospace]{cite}
\usepackage{threeparttable}

\begin{document}

\title{Bayesian Lower Bounds for Dense or Sparse (Outlier) Noise  in the RMT Framework}

\author{\IEEEauthorblockN{Virginie Ollier\IEEEauthorrefmark{1}\IEEEauthorrefmark{2},
R\'{e}my Boyer\IEEEauthorrefmark{2},
Mohammed Nabil El Korso\IEEEauthorrefmark{3} and
Pascal Larzabal\IEEEauthorrefmark{1}}
\IEEEauthorblockA{\IEEEauthorrefmark{1}SATIE, UMR 8029, Universit\'{e} Paris-Saclay, ENS Cachan, Cachan, France}
\IEEEauthorblockA{\IEEEauthorrefmark{2}L2S, UMR 8506, Universit\'{e} Paris-Saclay, Universit\'{e} Paris-Sud, Gif-sur-Yvette, France}
\IEEEauthorblockA{\IEEEauthorrefmark{3}LEME, EA 4416, Universit\'{e} Paris-Ouest, Ville d'Avray, France}
\thanks{This work was supported by the
following projects: MAGELLAN (ANR-14-CE23-0004-01)
and ICode blanc.}
}

\maketitle
\begin{abstract}
Robust estimation is an important and timely research subject. In this paper, we investigate performance lower bounds on the mean-square-error (MSE) of any estimator  for the Bayesian linear model, corrupted by a  noise distributed according to an i.i.d. Student's t-distribution. This class of prior parametrized by its degree of freedom is relevant to modelize either dense or sparse (accounting for outliers) noise. Using the hierarchical Normal-Gamma representation of  the Student's t-distribution, the Van Trees'  Bayesian Cram\'{e}r-Rao bound (BCRB) on the amplitude parameters is derived. Furthermore, the random matrix theory (RMT) framework is assumed, {\em i.e.}, the number of measurements and the number of unknown parameters grow jointly to infinity with an asymptotic finite  ratio.  Using some powerful results from the RMT, closed-form expressions of the BCRB are derived and studied. Finally,  we propose a  framework to  fairly compare two models corrupted by noises with different degrees of freedom for a  fixed common target signal-to-noise ratio (SNR). In particular, we focus our effort on the comparison of the BCRBs associated with two models corrupted by a sparse noise promoting outliers and  a dense (Gaussian) noise, respectively.
 \end{abstract}
 
\begin{IEEEkeywords}
Bayesian hierarchical linear model, Bayesian Cram\'{e}r-Rao bound, sparse outlier noise, dense noise, random matrix theory 
\end{IEEEkeywords}
 
\section{Introduction}

In the context of robust data modeling \cite{zoubir2012robust}, the measurement vector may be corrupted by noise containing outliers. This class of noise is sometimes referred to as sparse noise and is described by a distribution with heavy-tails \cite{mitra2010robust, mitra2010robust2, zhuang2014robust, newstadt2014robust, sundin2015combined, sundin2015bayesian}. Conversely, we usually call dense a noise that does not share this property and the most popular prior is probably Gaussian noise. Depending on the application context, outliers may be identified, \textit{e.g.}, as corrupted information or incomplete data \cite{luttinen2012bayesian}. 

A robust and relevant noise prior which is able to take into account outliers is the Student's t-distribution  with low degrees of freedom \cite{peel2000robust,kotz2004multivariate,christmas2014bayesian,zhang2014synthetic}. In addition, dense noise can also be encompassed thanks to the  Student's t-distribution prior for an infinite degree of freedom. A convenient framework to deal with a wide class of distributions is well known under the name of  
hierarchical Bayesian modeling.  The Bayesian hierarchical linear model (BHLM) with hierarchical noise prior is used in a wide range of applications, including fusion \cite{wei2014bayesian}, anomaly detection of hyperspectral images \cite{newstadt2014robust}, channel estimation \cite{pedersen2012application}, blind deconvolution \cite{kail2012blind}, segmentation of astronomical times series \cite{dobigeon2007joint}, \textit{etc}.

 In this work, we adopt such hierarchical prior framework due to its  flexibility and ability to modelize a wide class of priors. More precisely, the noise vector is assumed to follow  a circular {\rm i.i.d.} centered Gaussian prior with a variance defined by the inverse of an unknown  random hyper-parameter. In addition, if this hyper-parameter  is Gamma distributed  \cite{gelman2006prior,dahlin2012hierarchical}, then the marginalized joint pdf over the hyper-parameter is the  Student's t-distribution. 

The Van Trees' Bayesian Cram\'{e}r-Rao bound (${\rm BCRB}$) \cite{van2007bayesian} is a standard and fundamental lower bound on the mean-square-error (${\rm MSE}$) of any estimator. The aim of this work is to derive and analyze the ${\rm BCRB}$ of the amplitude parameters $(i)$ for  the considered noise prior and $(ii)$ using some powerful results from the random matrix theory (RMT) framework \cite{silverstein1995empirical,tulino2004random,couillet2011random}.  Regarding reference \cite{elkorso2016}, the proposed work is original in the sense that the noise prior is different and the asymptotic regime is assumed. Finally, note that reference \cite{prasad2013cramer} tackles a similar problem but does not assume the asymptotic context.

We use the following notation. Scalars, vectors and matrices are denoted by italic lower-case, boldface lower-case and boldface upper-case symbols, respectively. The symbol
 ${\rm Tr}[\cdot]$ stands for the trace operator.
The $K \times K$ identity matrix is denoted by ${\bf I}_K$ and $\mathbf{0}_{K \times 1}$ is the $K \times 1$ vector filled with zeros. The probability density function (pdf) of a given random variable $u$ is denoted by $p(u)$. The symbol $\mathcal{N} (\cdot,\cdot)$ refers to the Gaussian distribution, parametrized by its mean and covariance matrix, $\mathcal{G} (\cdot,\cdot)$ is the Gamma distribution, described by its shape and rate (inverse scale) parameters, while $\mathcal{IG} (\cdot,\cdot)$ is the inverse-Gamma distribution. If we have $ u \sim \mathcal{G} (a, b)$ then
 $p(u|a, b) = \frac{b^{a}u^{a-1} e^{-b u}} {\Gamma(a)}$, where $\Gamma(\cdot)$ is the Gamma function. And if $ u \sim \mathcal{IG} (a, b)$, then  $p(u|a, b) = \frac{b^{a}u^{-a-1} e^{-\frac{b}{u}}} {\Gamma(a)}$.  The non-standardized Student's t-distribution is defined by three parameters, through the pdf $p(u|\mu, \sigma^2, \nu,) = \frac{\Gamma(\frac{\nu+1}{2})}{\Gamma(\frac{\nu}{2})\sqrt{\pi \nu \sigma^2}}(1+\frac{1}{\nu}\frac{(u-\mu)^2}{\sigma^2})^{-\frac{\nu+1}{2}}$ such that $u \sim \mathcal{S}(\mu, \sigma^2, \nu).$ As regards the bivariate Normal-Gamma distribution, if we have 
$(u, w) \sim \rm{NormalGamma}(\mu, \lambda, a,b)$, then $p(u, w | \mu, \lambda, a,b ) = \frac{b^{a}\sqrt{\lambda}}{\Gamma(a)\sqrt{2\pi}}w^{a - \frac{1}{2}}e^{-b w}e^{-\frac{\lambda w (u-\mu)^2}{2}}$.
Finally, the symbol $\stackrel{a.s.} {\rightarrow}$ denotes almost sure convergence, $O(\cdot)$ is the big $O$ notation, $\lambda_i(\cdot)$ is the $i$-th eigenvalue of the considered matrix and
the symbol $\mathbb{E}_{{\bf
u}|{\bf
w}}$ refers to the expectation with respect to $p({\bf
u}|{\bf
w})$.

\section{Bayesian linear model corrupted by noise outliers}

\subsection{Definition of the random model }
 
Let ${\bf y}$ be the $N \times 1$ vector of measurements. The BHLM is defined by
\begin{equation}
\label{HBL}
{\bf y} = {\bf A}{\bf x}+ {\bf e},
\end{equation}
where each element $[\mathbf{A}]_{i,j}$ of the $N \times K$ matrix ${\bf A}$, with $K<N$, is drawn from an {\rm i.i.d.} as a single realization of a sub-Gaussian distribution with zero-mean and variance $1/N$ \cite{buldygin2000metric,couillet2011random}. The unknown amplitude vector is given by 
\begin{equation}
\label{modelx}
{\bf x} = [x_{1}, \hdots, x_{K}]^{T} \sim \mathcal{N} ({\bf
0}_{K \times 1},\sigma_{x}^{2} {\bf I}_K),
\end{equation}
where $\sigma_{x}^2$ is the known amplitude variance. In addition, the measurements are contaminated by a noise vector ${\bf e}$ which is assumed statistically independent from ${\bf x}$.

\subsection{Hierarchical Normal-Gamma representation}

The $i$-th noise sample  is assumed to be circular centered {\rm i.i.d.} Gaussian according to
\begin{equation}
\label{conde}
e_{i} | \gamma \sim \mathcal{N} \left(0,\frac{\sigma^2}{\gamma} \right),
\end{equation}
where $\frac{\gamma}{\sigma^2}$ is usually called the noise precision, $\gamma$ is an unknown hyper-parameter and $\sigma^2$ is a fixed scale parameter.

If the hyper-parameter is Gamma distributed according to 
\begin{equation}
\label{Gammm}
\gamma \sim \mathcal{G} \left(\frac{\nu}{2},\frac{\nu}{2}\right),
\end{equation}
where $\nu$ is the number of degrees of freedom, the joint distribution of $(e_{i}, \gamma)$ follows a  Normal-Gamma distribution \cite{bernardo2001bayesian} such as
\begin{equation}
(e_{i}, \gamma) \sim \rm{NormalGamma}\left(0, \frac{1}{\sigma^2}, \frac{\nu}{2}, \frac{\nu}{2}\right).
\end{equation}
The marginal distribution of the joint pdf over the hyper-parameter $\gamma$ leads to a non-standardized Student's t-distribution, given by \cite{svensen2005robust, christmas2014bayesian}
\begin{equation}
\mathcal{S}(e_{i} | 0, \sigma^2, \nu) = \int_{0
}^{\infty} \mathcal{N} \left(e_{i} |0,\frac{\sigma^2}{\gamma} \right)\mathcal{G} \left(\gamma | \frac{\nu}{2},\frac{\nu}{2}\right)\rm{d}\gamma,
\end{equation}
such that $e_{i} \sim \mathcal{S}(0, \sigma^2, \nu)$.

As $\nu \rightarrow \infty$, the distribution tends to a Gaussian  with zero-mean and variance $\sigma^2$, while it becomes more heavy-tailed when $\nu$ is small \cite{sfikas2007robust,zhang2014synthetic}.
With (\ref{conde}) and (\ref{Gammm}), and knowing that $\frac{1}{\gamma}\sim \mathcal{IG} (\frac{\nu}{2}, \frac{\nu}{2})$ , we notice that the variance, noted $\sigma_e^2$ of each noise entry of ${\bf e}$, is given by the following expression
\begin{equation}
\sigma_e^2 = \mathbb{E}_{\gamma}\mathbb{E}_{e_{i} | \gamma}\left\{e_{i}^2\right\} =
\sigma^2\mathbb{E}_{\gamma}\left\{\frac{1}{\gamma}\right\}=
\sigma^2 \frac{\nu}{\nu-2},
\end{equation}
in which $\nu >2$.

\section{${\rm BCRB}$  for Student's t-distribution}

The vector of unknown parameters, denoted by $\boldsymbol{\theta}$, encompasses the amplitude vector and the noise hyper-parameter, {\em i.e.},
\begin{equation}
\boldsymbol{\theta}=[\mathbf{x}^T,\gamma]^{T}.
\end{equation}
Given an independence assumption between $\mathbf{x}$ and $\gamma$, the joint pdf $p({\bf y},\boldsymbol{\theta})$ can be decomposed as
\begin{equation}
p({\bf y},\boldsymbol{\theta}) = 
p({\bf y}|
\boldsymbol{\theta})p(\boldsymbol{\theta}) = p({\bf y}|
\boldsymbol{\theta})p(\mathbf{x})p(\gamma).
\end{equation}

Let us note $\boldsymbol{\hat{\theta}}$ an estimator of the unknown vector $\boldsymbol{\theta}$. Then, the mean square error (${\rm MSE}$), directly linked to the error covariance matrix, verifies the following inequality 
\begin{equation}
\label{MSE}
 {\rm MSE} (\boldsymbol{\theta}) = {\rm Tr}\left[\mathbb{E}_{{\bf y},\boldsymbol{\theta}}\left\{   (\boldsymbol{\theta}-\boldsymbol{\hat{\theta}})(\boldsymbol{\theta}-\boldsymbol{\hat{\theta}})^T  \right\}\right]  \geq {\rm Tr}\left[{\bf C}\right],
\end{equation}
where  $\mathbf{C}$ is the $(K+1) \times (K+1)$ ${\rm BCRB}$ matrix defined as the inverse of the Bayesian Information Matrix (BIM) $\mathbf{J}$. We can show that the BIM has a block-diagonal structure due to the independence between parameters. Thus, we write
\begin{equation}
\mathbf{J} = \begin{bmatrix}
    \mathbf{J}_{\mathbf{x},\mathbf{x}} &  \mathbf{0}_{K \times 1}\\
    \mathbf{0}_{1 \times K} & \mathrm{J}_{\gamma,\gamma}
  \end{bmatrix}.
\end{equation}
We assume an identifiable BHLM model so that, under weak regularity conditions \cite{van2007bayesian}, the BIM is given by
\begin{eqnarray}
{\bf J} = \mathbb{E}_{\boldsymbol{\theta}}\left\{ {\bf J}_D^{(\boldsymbol{\theta},\boldsymbol{\theta} )}\right\} +  {\bf J}_P^{(\boldsymbol{\theta},\boldsymbol{\theta} )}   + {\bf J}_{HP}^{(\boldsymbol{\theta},\boldsymbol{\theta} )},
\end{eqnarray}
in which 
\begin{equation}
[{\bf J}_D^{(\boldsymbol{\theta},\boldsymbol{\theta} )}]_{i,j} = \mathbb{E}_{{\bf
y}|\boldsymbol{\theta}}
 \left\{ - \frac{\partial^2 \log p({\bf y} | \boldsymbol{\theta}) }{\partial \theta_i \partial \theta_j}\right\},
 \end{equation}
\begin{equation}
[{\bf J}_P^{(\boldsymbol{\theta},\boldsymbol{\theta} )}]_{i,j} = \mathbb{E}_{\mathbf{x}}
 \left\{ - \frac{\partial^2 \log p(\mathbf{x}) }{\partial \theta_i \partial \theta_j}\right\},
 \end{equation}
\begin{equation}
[{\bf J}_{HP}^{(\boldsymbol{\theta},\boldsymbol{\theta} )}]_{i,j} = \mathbb{E}_{\gamma}  \left\{ -
\frac{\partial^2 \log p(\gamma) }{\partial \theta_i \partial
\theta_j}\right\}
\end{equation}
for $(i,j) \in \left\{1,\dots,K+1\right\}^2$, and where ${\bf J}_D^{(\boldsymbol{\theta},\boldsymbol{\theta} )}$ is the Fisher Information Matrix (FIM) on $\boldsymbol{\theta}$, ${\bf J}_P^{(\boldsymbol{\theta},\boldsymbol{\theta} )}$ is the prior part of the BIM and ${\bf J}_{HP}^{(\boldsymbol{\theta},\boldsymbol{\theta} )}$ is the hyper-prior part.

 Correspondingly, we have
\begin{equation}
\label{BCRB}
\mathbf{C} = \mathbf{J} ^{-1} = \begin{bmatrix}
    \mathbf{C}_{\mathbf{x},\mathbf{x}} &  \mathbf{0}_{K \times 1}\\
    \mathbf{0}_{1 \times K} & \mathrm{C}_{\gamma,\gamma}
  \end{bmatrix}.
\end{equation}

Conditionally to $\boldsymbol{\theta}$, the observation vector $\mathbf{y}$ has the following Gaussian distribution 
\begin{equation}
{\bf y} | \boldsymbol{\theta} \sim \mathcal{N}\Big(\boldsymbol{\mu},\mathbf{R}\Big),
\end{equation}
where $\boldsymbol{\mu} = {\bf A}{\bf x}$ and $\mathbf{R} = \frac{ (\nu -2)\sigma_e^2}{\nu \gamma}  {\bf I}_{N}$.
In what follows, we directly make use of the Slepian-Bangs formula \cite[p. 378]{stoica2005spectral}
\begin{equation}
[{\bf J}_D^{(\boldsymbol{\theta},\boldsymbol{\theta} )}]_{i,j} = \left(\frac{\partial \boldsymbol{\mu}}{\partial \theta_{i}}\right)^{T}\mathbf{R}^{-1}\frac{\partial \boldsymbol{\mu}}{\partial \theta_{j}} + \frac{1}{2}{\rm Tr}\left[\frac{\mathbf{R}}{\partial \theta_{i}}\mathbf{R}^{-1}\frac{\mathbf{R}}{\partial \theta_{j}}\mathbf{R}^{-1}\right].
\end{equation}
 This leads to 
\begin{equation}
{\bf J}_D^{(\mathbf{x},\mathbf{x} )}=\frac{\nu \gamma}{(\nu -2)\sigma_e^2}{\bf A}^T {\bf A}.
\end{equation}
Using the fact that $\mathbf{R}^{-1} = \frac{\gamma}{\sigma^2}{\bf I}_{N}$, we obtain
\begin{equation}
 {\rm J}_D^{(\gamma,\gamma )}=\frac{\sigma^4}{2 \gamma^{4}}{\rm Tr}\Big[\mathbf{R}^{-2}\Big] = \frac{N }{2 \gamma^{2}}.
\end{equation}
According to (\ref{modelx}) and considering independent amplitudes, we have
\begin{equation}
-\log p(\mathbf{x} ) =
\sum_{i=1}^{K}\left(\frac{1}{2}\log(2\pi\sigma_x^2)+
 \frac{x_i^2}{2\sigma_x^2}\right).
\end{equation}
Consequently,
\begin{equation}
{\bf J}_P^{(\mathbf{x},\mathbf{x} )}=\frac{1}{\sigma_x^2}{\bf I}_{K}.
\end{equation}

The BIM $\mathbf{J}$ is therefore composed of the following terms:
\begin{equation} \mathbf{J}_{\mathbf{x},\mathbf{x}} = \mathbb{E}_{\gamma}\left\{{\bf J}_D^{(\mathbf{x},\mathbf{x} )}\right\}+{\bf J}_P^{(\mathbf{x},\mathbf{x} )},
\end{equation} 
\begin{equation}
{\rm J}_{\gamma,\gamma}  = \mathbb{E}_{\gamma}\left\{ {\rm J}_D^{(\gamma,\gamma )}\right\}+{\rm J}_{HP}^{(\gamma,\gamma )}.
\end{equation}
The hyper-prior part of the BIM is given by 
\begin{equation}
{\rm J}_{HP}^{(\gamma,\gamma )}=\mathbb{E}_{\gamma} 
 \left\{ -
\frac{\partial^2 \log p(\gamma) }{\partial \gamma^{2}}\right\}  =\frac{\nu-2}{2}
\mathbb{E}_{\gamma}\left\{\frac{1}{\gamma^{2}}\right\}.
\end{equation}
The second-order moment of an inverse-Gamma distributed random variable is given by 
\begin{equation}
\label{moy_inv_carre}
\mathbb{E}_{\gamma}\left\{\frac{1}{\gamma^{2}}\right\} 
=\frac{\nu^2}{(\nu-2)(\nu-4)},
\end{equation}
where $\nu > 4$. This finally leads to
\begin{equation}\label{Jgamma}
{\rm J}_{\gamma,\gamma}  = \frac{N\nu^2}{2(\nu-2)(\nu-4)} + \frac{\nu^2}{2(\nu-4)}.
\end{equation}

Inverting the BIM, we obtain the ${\rm BCRB}$  for the amplitude parameters
\begin{align}
\label{BCRB_x_fin}
{\rm BCRB}(\mathbf{x})=\frac{{\rm Tr}\left[\mathbf{C}_{\mathbf{x},\mathbf{x}}   \right]}{K}\ \ \mbox{with} \  
\mathbf{C}_{\mathbf{x},\mathbf{x}}   = \sigma_x^2\left(  r {\bf A}^T {\bf A}+{\bf I}_{K}  \right)^{-1},
\end{align}
where   $r={\rm SNR}\frac{\nu }{\nu -2}$ with ${\rm SNR} =\frac{\sigma_x^2}{\sigma_e^2}$ (signal-to-noise ratio).


\section{${\rm BCRB}$ in the asymptotic framework}


\subsection{RMT framework}
\label{sec : RMT}

In this section, we consider the context of large random matrices, {\em i.e.}, for $K, N \rightarrow \infty$ with $\frac{K}{N} \rightarrow \beta \in (0,1)$. The derived ${\rm BCRB}$ in this context is the asymptotic normalized ${\rm BCRB}$ defined by
\begin{equation}
{\rm BCRB}(\mathbf{x}) \stackrel{a.s.} {\rightarrow} {\rm BCRB}^{\infty}(\mathbf{x}).
\end{equation}

Using (\ref{BCRB_x_fin}) with \cite [p. 11]{tulino2004random}, we obtain
\begin{equation}
\label{BCRB_x_RMT}
{\rm BCRB}^{\infty}(\mathbf{x})=\sigma_x^2 \left(1-\frac{f( r,\beta)}{4 r \beta}\right)
\end{equation}
 and $f( r,\beta) = \left(\sqrt{r(1+\sqrt{\beta})^2 +1}- \sqrt{r(1-\sqrt{\beta})^2 +1}\right)^2$.

\subsection{Limit analytical expressions}
\label{sec : asymptotic}


\begin{itemize}
\item For $\beta \ll 1$, {\em i.e.}, $ K \ll N$, after some manipulations and discarding the terms of order superior or equal to $O(\beta^2)$, we obtain
\begin{equation}
f(r,\beta) \thickapprox \frac{4\beta r^2}{r+1}.
\end{equation}
Therefore, an asymptotic analytical expression of the ${\rm BCRB}$, in the RMT framework, is given by
\begin{equation}
\label{BCRB_x_beta_low}
{\rm BCRB}^{\infty}(\mathbf{x}) \approx \frac{\sigma_x^2}{r+1} = \frac{(\nu-2)\sigma_x^2}{\nu(1+{\rm SNR})-2}.
\end{equation}

%
%

\item For small $r$, also meaning small ${\rm SNR}$, according to the Neumann series expansion \cite{arfken2005mathematical}, we have $\left(r {\bf A}^T{\bf A} + {\bf I}_K\right)^{-1} \approx {\bf I}_K -  r {\bf A}^T{\bf A}$ if the maximal eigenvalue $\lambda_{\max}(r {\bf A}^T{\bf A}) <1$. Observe that $r\lambda_{\max}( {\bf A}^T{\bf A}) \stackrel{a.s.} {\rightarrow} r(1+\sqrt{\beta})^2$ \cite{silverstein1995empirical,tulino2004random,couillet2011random}. In addition, if ${\rm SNR}$ is sufficiently small with respect to $(\nu-2)/(4\nu)$ then
\begin{align}
\label{BCRB_x_r_low}
\nonumber
{\rm BCRB}(\mathbf{x})  & \approx \frac{\sigma_x^2}{K}\left({\rm Tr}\left[{\bf I}_{K} \right] - r {\rm Tr}\left[{\bf A}^T {\bf A} \right]  \right) \\ & \stackrel{a.s.}
 {\rightarrow} \sigma_x^2(1-r)
 = \frac{\sigma_x^2}{\nu-2}(\nu-2-\nu {\rm SNR}).
\end{align}

\item For large $r$, also meaning large ${\rm SNR}$, we have
\begin{align}
\label{BCRB_x_r_large}
\nonumber {\rm BCRB}(\mathbf{x})&\approx \frac{\sigma_x^2}{rK}\left({\rm Tr}\left[\left( {\bf A}^T {\bf A}\right)^{-1} \right]-\frac{1}{r}{\rm Tr}\left[\left( {\bf A}^T {\bf A}\right)^{-2} \right] \right)\\
& \nonumber  \stackrel{a.s.} {\rightarrow}  \frac{\sigma_x^2}{r} \left(\frac{1}{1-\beta} -\frac{1}{r} \frac{1}{(1-\beta)^3} \right)\\
&
= \frac{(\nu-2)\sigma_x^2}{\nu{\rm SNR}(1-\beta)}\left(1-\frac{\nu-2}{\nu {\rm SNR}(1-\beta)^2} \right),
\end{align}
since \cite{silverstein1995empirical,tulino2004random,couillet2011random}
\begin{align}
\frac{1}{K}{\rm Tr}\left[\left( {\bf A}^T {\bf A}\right)^{-1} \right] &\stackrel{a.s.} {\rightarrow} \frac{1}{1-\beta},\\
\frac{1}{K}{\rm Tr}\left[\left( {\bf A}^T {\bf A}\right)^{-2} \right] &\stackrel{a.s.} {\rightarrow} \frac{1}{(1-\beta)^3}.
\end{align}
\end{itemize}
\subsection{Comparison between two models with a target common $\rm SNR$}
\label{sec : comparison}

We consider two different models:
\begin{align}
(M_{0})&: \ {\bf y}_{0} = {\bf A}{\bf x}+ {\bf e}_{0} \ \ {\rm with} \  e_{i_{0}} \sim \mathcal{S}(0, \sigma_{0}^2, \nu_{0}),\\
(M_{1})&: \ {\bf y}_{1} = {\bf A}{\bf x}+ {\bf e}_{1} \ \ {\rm with} \  e_{i_{1}} \sim \mathcal{S}(0, \sigma_{1}^2, \nu_{1}).
\end{align}

Model $(M_{0})$ is the reference model and model $(M_{1})$ is the alternative one. According to (\ref{BCRB_x_RMT}), the asymptotic normalized ${\rm BCRB}$ for the $k$-th model with $k\in \{0,1\}$ is defined by
\begin{equation}
{\rm BCRB}_k^{\infty}(\mathbf{x})=\sigma_x^2 \left(1-\frac{f( r_k,\beta)}{4 r_k \beta}\right)
\end{equation}
where $r_k = {\rm SNR}_{k} \frac{\nu_k}{\nu_k-2}$ with  $ {\rm SNR}_{k} = \frac{\sigma_{x}^2}{\sigma_{e_{k}}^2}$. 
A fair methodology to compare the bounds  ${\rm BCRB}_{0}(\mathbf{x})$ and ${\rm BCRB}_{1}(\mathbf{x})$ is to impose a common target $\rm SNR$ for the models $(M_{0})$ and $(M_{1})$, {\em i.e.}, ${\rm SNR}_{0} = {\rm SNR}_{1}$. A simple derivation shows that to reach the target $\rm SNR$, we must have  $r_1 = \frac{\nu_1(\nu_0-2)}{\nu_0(\nu_1-2)}r_0$. Specifically,  the corresponding ${\rm BCRBs}$ are the following ones:
\begin{align}
{\rm BCRB}_0^{\infty}(\mathbf{x})&=\sigma_x^2 \left(1-\frac{f( r_0,\beta)}{4 r_0 \beta}\right),\\ \label{BCRB_1}
{\rm BCRB}_{1}^{\infty}(\mathbf{x})   & =\sigma_x^2 \left(1-\frac{\nu_0(\nu_1-2) f( \frac{\nu_1(\nu_0-2)}{\nu_0(\nu_1-2)}r_0,\beta)}{4 \nu_1(\nu_0-2)r_0 \beta}\right).
\end{align}



Recall that the  Student's t-distribution is well known to promote noise outliers thanks to its heavy-tails property unlike the Gaussian distribution. So, an interesting scenario arises  when $\nu_{1} \rightarrow \infty$. In this case, the Student's t-distribution converges to the Gaussian one \cite{kotz2004multivariate} and  (\ref{BCRB_1}) tends to
\begin{equation}
\label{approximate}
{\rm BCRB}_{1}^{\infty}(\mathbf{x})   \overset{\nu_{1} \rightarrow \infty} = \sigma_x^2 \left(1-   \frac{\nu_0f\Big( \frac{\nu_0-2}{\nu_0} r_0,\beta\Big)}{4 (\nu_0-2)r_0 \beta}\right).
\end{equation}


\subsection{Numerical simulations}
\label{sec : simus}

\begin{figure}[tb] 
  \centering
  \centerline{\includegraphics[width=8cm, height=4.5cm]{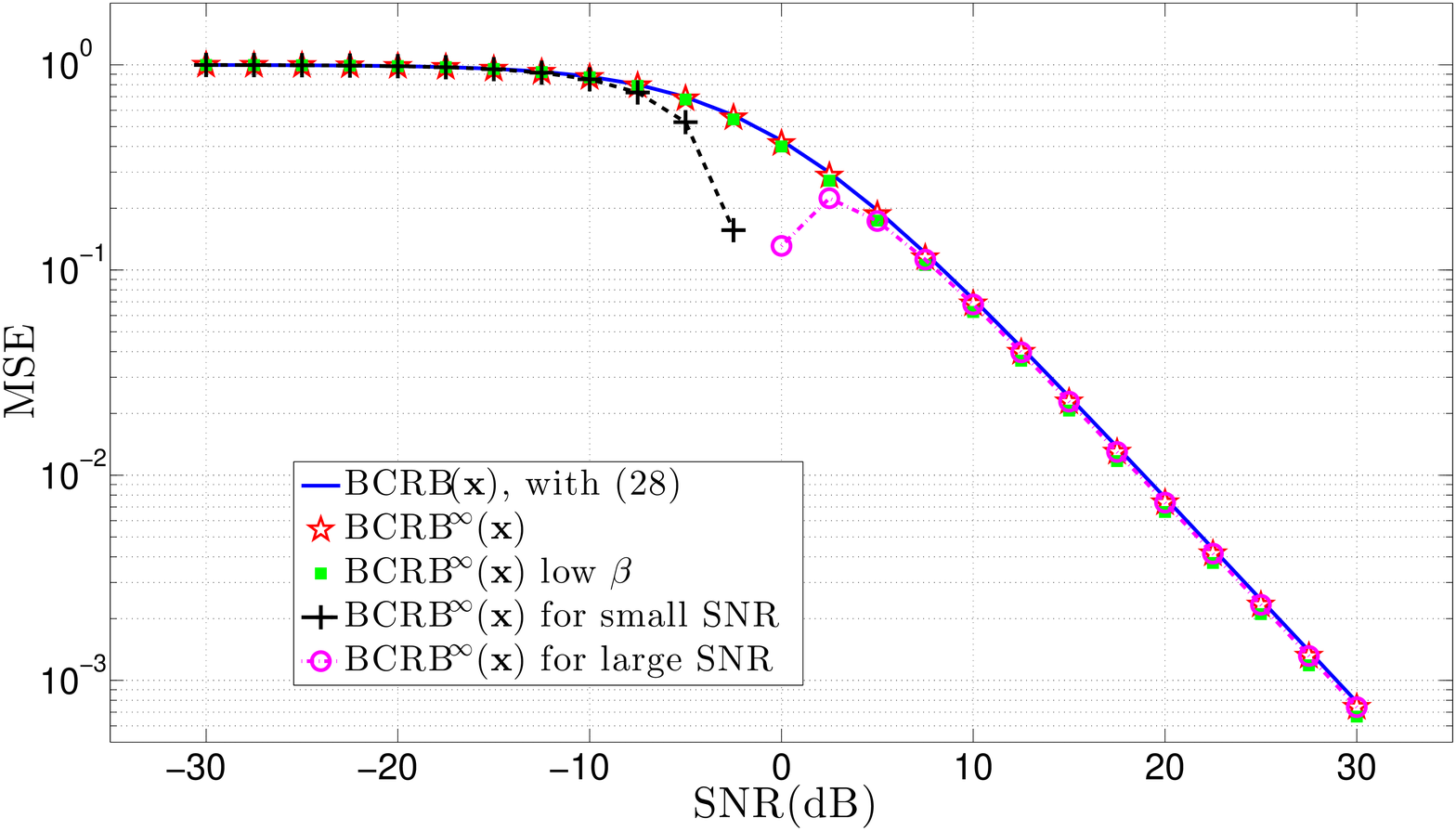}}
\caption{${\rm BCRB}$($\mathbf{x}$) as a function of ${\rm SNR}$ in dB with specific limit approximations, in the RMT framework.} \label{Fig.:res1}
\end{figure}

In the following simulations, we consider $N=100$ and $K=10$ so that $\beta \ll 1$. The amplitude variance $\sigma_x^2$ is fixed to $1$.
In Fig. \ref{Fig.:res1}, we plot the ${\rm BCRB}$ of the amplitude vector $\mathbf{x}$, as defined by equations (\ref{BCRB_x_fin}) and (\ref{BCRB_x_RMT}) (asymptotic expression), (\ref{BCRB_x_beta_low}) (small $\beta$), (\ref{BCRB_x_r_low}) (small ${\rm SNR}$) and (\ref{BCRB_x_r_large}) (large ${\rm SNR}$), as a function of the ${\rm SNR}$ in dB for $\nu=6$.

We notice that ${\rm BCRB}(\mathbf{x})$ coincides precisely with its asymptotic expression in (\ref{BCRB_x_RMT}). Thus, the RMT framework predicts precisely the behavior of the ${\rm BCRB}$ of the amplitude as $K, N \rightarrow \infty$ with $\frac{K}{N} \rightarrow \beta$ and allows us to obtain a closed-form expression. Such limit remains correct even for values of $N$ and $K$ that are relatively not quite large. The expression of the ${\rm BCRB}$ obtained with (\ref{BCRB_x_beta_low}) is a good approximation since here, we have $\beta = 0.1 \ll 1$. Finally, we notice that the curves obtained for low and high $\rm SNR$ approximate very well the ${\rm BCRB}$ of the amplitude, asymptotically.

\begin{figure}[tb] 
  \centering
  \centerline{\includegraphics[width=8cm, height=4.5cm]{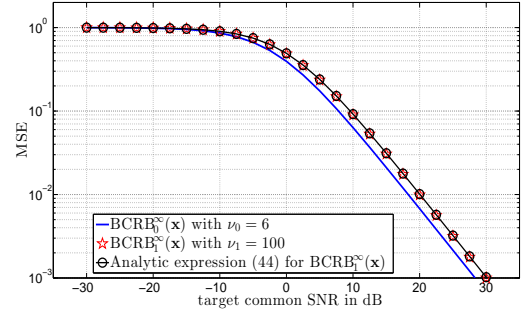}}
\caption{Asymptotic normalized ${\rm BCRBs}$ for models $(M_{0})$ and $(M_{1})$ {\em vs.} a common $\rm SNR$} \label{Fig.:res2}
\end{figure}


In Fig. \ref{Fig.:res2}, as exposed in section \ref{sec : comparison}, we consider two different models, with a different value for the number of degrees of freedom $\nu$. We notice that a lower performance bound is achieved with $\nu_{0}=6$, especially in the low noise regime, than with  $\nu_{1}=100$. Furthermore, the approximation in (\ref{approximate}) is correct, since $\nu_{1}$ has a large value. A low value for the number of degrees of freedom is well-adapted for the modelization of sparse (outlier) noise, characterized by a heavy-tailed distribution \cite{tzagkarakis2009bayesian, amini2011compressibility}. This large level in heavy-tailedness leads to robustness \cite{lange1989robust,delannay2008improving, zoubir2012robust} while a Gaussian noise model (large degree of freedom) corresponds to a dense noise type. Thus, we can hope to achieve better estimation performances if we consider a model, which promotes sparsity and the presence of outliers in data.
%

\section{Conclusion}

This work discusses fundamental Bayesian lower bounds for multi-parameter robust estimation. More precisely, we consider a Bayesian linear model corrupted by a sparse noise following a Student's t-distribution. This class of prior can efficiently modelize outliers. Using the hierarchical Normal-Gamma representation of the Student's t-distribution, the Van Trees' Bayesian lower bound (${\rm BCRB}$) is derived   for unknown amplitude parameters in an asymptotic context. By asymptotic, it means that the number of measurements and the number of unknown parameters grow to infinity at a finite rate. Consequently,  closed-form expressions of the ${\rm BCRB}$ are obtained using some powerful results from the large random matrix theory. Finally, a  framework is provided to fairly compare two models corrupted by noises with different degrees of freedom for a fixed common target $\rm SNR$. We recall that a small degree of freedom promotes outliers in the sense that the noise prior has heavy-tails. For the amplitude, a lower performance bound is achieved when the number of degrees of freedom is small.

%



 \bibliographystyle{IEEEtran}
\bibliography{nab}

\begin{thebibliography}{10}
\providecommand{\url}[1]{#1}
\csname url@samestyle\endcsname
\providecommand{\newblock}{\relax}
\providecommand{\bibinfo}[2]{#2}
\providecommand{\BIBentrySTDinterwordspacing}{\spaceskip=0pt\relax}
\providecommand{\BIBentryALTinterwordstretchfactor}{4}
\providecommand{\BIBentryALTinterwordspacing}{\spaceskip=\fontdimen2\font plus
\BIBentryALTinterwordstretchfactor\fontdimen3\font minus
  \fontdimen4\font\relax}
\providecommand{\BIBforeignlanguage}[2]{{%
\expandafter\ifx\csname l@#1\endcsname\relax
\typeout{** WARNING: IEEEtran.bst: No hyphenation pattern has been}%
\typeout{** loaded for the language `#1'. Using the pattern for}%
\typeout{** the default language instead.}%
\else
\language=\csname l@#1\endcsname
\fi
#2}}
\providecommand{\BIBdecl}{\relax}
\BIBdecl

\bibitem{zoubir2012robust}
A.~M. Zoubir, V.~Koivunen, Y.~Chakhchoukh, and M.~Muma, ``Robust estimation in
  signal processing: A tutorial-style treatment of fundamental concepts,''
  \emph{IEEE Signal Processing Magazine}, vol.~29, no.~4, pp. 61--80, 2012.

\bibitem{mitra2010robust}
K.~Mitra, A.~Veeraraghavan, and R.~Chellappa, ``Robust{ RVM} regression using
  sparse outlier model,'' in \emph{IEEE Conference on Computer Vision and
  Pattern Recognition (CVPR)}, San Francisco, CA, 2010, pp. 1887--1894.

\bibitem{mitra2010robust2}
------, ``Robust regression using sparse learning for high dimensional
  parameter estimation problems,'' in \emph{IEEE Int. Conf. Acoust., Speech and
  Signal Processing (ICASSP)}, Dallas, TX, 2010, pp. 3846--3849.

\bibitem{zhuang2014robust}
P.~Zhuang, W.~Wang, D.~Zeng, and X.~Ding, ``Robust mixed noise removal with
  non-parametric {B}ayesian sparse outlier model,'' in \emph{16th International
  Workshop on Multimedia Signal Processing (MMSP)}, Jakarta, Indonesia, 2014,
  pp. 1--5.

\bibitem{newstadt2014robust}
G.~E. Newstadt, A.~O. Hero, and J.~Simmons, ``Robust spectral unmixing for
  anomaly detection.'' in \emph{IEEE Workshop on Statistical Signal Processing
  (SSP)}, Gold Coast, VIC, 2014, pp. 109--112.

\bibitem{sundin2015combined}
M.~Sundin, S.~Chatterjee, and M.~Jansson, ``Combined modeling of sparse and
  dense noise improves {B}ayesian {RVM},'' in \emph{22nd European Signal
  Processing Conference (EUSIPCO)}, Lisbon, Portugal, 2014, pp. 1841--1845.

\bibitem{sundin2015bayesian}
------, ``Bayesian learning for robust {P}rincipal {C}omponent {A}nalysis,'' in
  \emph{23rd European Signal Processing Conference (EUSIPCO)}, Nice, France,
  2015, pp. 2361--2365.

\bibitem{luttinen2012bayesian}
J.~Luttinen, A.~Ilin, and J.~Karhunen, ``Bayesian robust {PCA} of incomplete
  data,'' \emph{Neural Processing Letters}, vol.~36, no.~2, pp. 189--202, 2012.

\bibitem{peel2000robust}
D.~Peel and G.~J. McLachlan, ``Robust mixture modelling using the t
  distribution,'' \emph{Statistics and computing}, vol.~10, no.~4, pp.
  339--348, 2000.

\bibitem{kotz2004multivariate}
S.~Kotz and S.~Nadarajah, \emph{Multivariate t-distributions and their
  applications}.\hskip 1em plus 0.5em minus 0.4em\relax Cambridge University
  Press, 2004.

\bibitem{christmas2014bayesian}
J.~Christmas, ``Bayesian spectral analysis with {S}tudent-t noise,'' \emph{IEEE
  Transactions on Signal Processing}, vol.~62, no.~11, pp. 2871--2878, 2014.

\bibitem{zhang2014synthetic}
H.~Zhang, Q.~M.~J. Wu, T.~M. Nguyen, and X.~Sun, ``Synthetic aperture radar
  image segmentation by modified {S}tudent's t-mixture model,'' \emph{IEEE
  Transactions on Geoscience and Remote Sensing}, vol.~52, no.~7, pp.
  4391--4403, 2014.

\bibitem{wei2014bayesian}
Q.~Wei, N.~Dobigeon, and J.-Y. Tourneret, ``Bayesian fusion of hyperspectral
  and multispectral images,'' in \emph{IEEE Int. Conf. on Acoust., Speech and
  Signal Processing (ICASSP)}, Florence, Italy, 2014, pp. 3176--3180.

\bibitem{pedersen2012application}
N.~L. Pedersen, C.~N. Manch{\'o}n, D.~Shutin, and B.~H. Fleury, ``Application
  of {B}ayesian hierarchical prior modeling to sparse channel estimation,'' in
  \emph{IEEE International Conference on Communications (ICC)}, Ottawa, ON,
  2012, pp. 3487--3492.

\bibitem{kail2012blind}
G.~Kail, J.-Y. Tourneret, F.~Hlawatsch, and N.~Dobigeon, ``Blind deconvolution
  of sparse pulse sequences under a minimum distance constraint: A partially
  collapsed {G}ibbs sampler method,'' \emph{IEEE Transactions on Signal
  Processing}, vol.~60, no.~6, pp. 2727--2743, 2012.

\bibitem{dobigeon2007joint}
N.~Dobigeon, J.-Y. Tourneret, and J.~D. Scargle, ``Joint segmentation of
  multivariate astronomical time series: Bayesian sampling with a hierarchical
  model,'' \emph{IEEE Transactions on Signal Processing}, vol.~55, no.~2, pp.
  414--423, 2007.

\bibitem{gelman2006prior}
A.~Gelman, ``Prior distributions for variance parameters in hierarchical models
  (comment on article by {B}rowne and {D}raper),'' \emph{Bayesian Analysis},
  vol.~1, no.~3, pp. 515--534, 2006.

\bibitem{dahlin2012hierarchical}
J.~Dahlin, F.~Lindsten, T.~B. Sch{\"o}n, and A.~Wills, ``Hierarchical
  {B}ayesian {ARX} models for robust inference,'' in \emph{16th IFAC Symposium
  on System Identification (SYSID)}, Brussels, Belgium, 2012, pp. 131--136.

\bibitem{van2007bayesian}
H.~L. Van~Trees and K.~L. Bell, \emph{Bayesian bounds for parameter estimation
  and nonlinear filtering/tracking}.\hskip 1em plus 0.5em minus 0.4em\relax New
  York: Wiley-IEEE Press, 2007.

\bibitem{silverstein1995empirical}
J.~W. Silverstein and Z.~Bai, ``On the empirical distribution of eigenvalues of
  a class of large dimensional random matrices,'' \emph{Journal of Multivariate
  analysis}, vol.~54, no.~2, pp. 175--192, 1995.

\bibitem{tulino2004random}
A.~M. Tulino and S.~Verd{\'u}, \emph{Random matrix theory and wireless
  communications}.\hskip 1em plus 0.5em minus 0.4em\relax Foundations and
  Trends in Communications and Information Theory. Now Publishers Inc., 2004,
  vol.~1, no.~1.

\bibitem{couillet2011random}
R.~Couillet and M.~Debbah, \emph{Random matrix methods for wireless
  communications}.\hskip 1em plus 0.5em minus 0.4em\relax Cambridge University
  Press, 2011.

\bibitem{elkorso2016}
M.~N. El~Korso, R.~Boyer, P.~Larzabal, and B.-H. Fleury, ``Estimation
  performance for the {B}ayesian hierarchical linear model,'' \emph{IEEE Signal
  Processing Letters}, vol.~23, no.~4, pp. 488--492, 2016.

\bibitem{prasad2013cramer}
R.~Prasad and C.~R. Murthy, ``Cram{\'e}r-{R}ao-type bounds for sparse
  {B}ayesian learning,'' \emph{IEEE Transactions on Signal Processing},
  vol.~61, no.~3, pp. 622--632, 2013.

\bibitem{buldygin2000metric}
V.~V. Buldygin and Y.~V. Kozachenko, \emph{Metric characterization of random
  variables and random processes}.\hskip 1em plus 0.5em minus 0.4em\relax
  American Mathematical Soc., 2000, vol. 188.

\bibitem{bernardo2001bayesian}
J.~M. Bernardo and A.~F.~M. Smith, \emph{Bayesian theory}.\hskip 1em plus 0.5em
  minus 0.4em\relax New York: J. Wiley, 1994.

\bibitem{svensen2005robust}
M.~Svens{\'e}n and C.~M. Bishop, ``Robust {B}ayesian mixture modelling,''
  \emph{Neurocomputing}, vol.~64, pp. 235--252, 2005.

\bibitem{sfikas2007robust}
G.~Sfikas, C.~Nikou, and N.~Galatsanos, ``Robust image segmentation with
  mixtures of {S}tudent's t-distributions,'' in \emph{International Conference
  on Image Processing (ICIP)}, vol.~1, Sant Antonio, TX, 2007, pp. I--273 --
  I--276.

\bibitem{stoica2005spectral}
P.~Stoica and R.~L. Moses, \emph{Spectral analysis of signals}.\hskip 1em plus
  0.5em minus 0.4em\relax Pearson Prentice Hall, Upper Saddle River, NJ, 2005.

\bibitem{arfken2005mathematical}
G.~B. Arfken and H.~J. Weber, \emph{Mathematical methods for physicists, sixth
  edition}.\hskip 1em plus 0.5em minus 0.4em\relax Academic press, 2005.

\bibitem{tzagkarakis2009bayesian}
G.~Tzagkarakis and P.~Tsakalides, ``Bayesian compressed sensing of a highly
  impulsive signal in heavy-tailed noise using a multivariate {C}auchy prior,''
  in \emph{17th European Signal Processing Conference}, Glasgow, Scotland,
  2009, pp. 2293--2297.

\bibitem{amini2011compressibility}
A.~Amini, M.~Unser, and F.~Marvasti, ``Compressibility of deterministic and
  random infinite sequences,'' \emph{IEEE Transactions on Signal Processing},
  vol.~59, no.~11, pp. 5193--5201, 2011.

\bibitem{lange1989robust}
K.~L. Lange, R.~J.~A. Little, and J.~M.~G. Taylor, ``Robust statistical
  modeling using the t distribution,'' \emph{Journal of the American
  Statistical Association}, vol.~84, no. 408, pp. 881--896, 1989.

\bibitem{delannay2008improving}
N.~Delannay, C.~Archambeau, and M.~Verleysen, ``Improving the robustness to
  outliers of mixtures of probabilistic {PCA}s,'' \emph{Advances in Knowledge
  Discovery and Data Mining}, vol. 5012, pp. 527--535, 2008.

\end{thebibliography}

\end{document}